\documentclass[10pt,conference]{IEEEtran}
\IEEEoverridecommandlockouts
\usepackage{cite}
\usepackage{caption}

\usepackage{placeins} 

\usepackage{cuted}

\usepackage{float}
\usepackage{amsmath,amssymb,amsfonts}
\usepackage{algorithm}
\usepackage{algorithmicx}
\usepackage{algpseudocode}
\usepackage{graphicx}
\usepackage{textcomp}
\usepackage{xcolor}
\usepackage[left=0.76in, right=0.65in, top=0.76in, bottom=1.05in]{geometry}

\title{Coupling-Aware Pinching-Antenna Radiation for Heterogeneous NOMA Users}
\author{\IEEEauthorblockN{Ishtiaque Ahmed, Leila Musavian, and Haris Pervaiz}

 \thanks{The authors are with the School of Computer Science and Electronic Engineering, University of Essex, Wivenhoe Park, Colchester CO4 3SQ, United Kingdom (e-mail: { \{ishtiaque.ahmed, leila.musavian, haris.pervaiz\}@essex.ac.uk).} This work has been submitted to
the IEEE for possible publication. Copyright may be transferred
without notice, after which this version may no longer be accessible.

This work was supported by the UK Research and Innovation under the UK government’s
Horizon Europe funding guarantee through MSCA-DN SCION Project Grant
Agreement No.101072375 [grant number: EP/X027201/1].}
 }
    
\begin{document}
\maketitle

\begin{abstract}
Pinching-antenna systems (PASS) offer reconfigurable wireless channels via low-cost dielectric mediums by creating line-of-sight (LoS) communication links. Most of the existing studies on PASS consider equal-power radiation from pinching antennas for conventional bit-based communication, whereas flexible radiation remains largely unexplored, particularly for heterogeneous semantic and bit users. In this paper, we investigate the performance of semantic communication (SC) using an adjustable waveguide-antenna spacing mechanism over PASS for the users power-allocation coefficients and antenna positions optimization. Specifically, the non-orthogonal multiple access (NOMA)-assisted heterogeneous users are served by multiple pinching antennas using spacing-aware adjustable radiation ratios. Under this setting, we maximize the semantic spectral efficiency (SE) subject to the bit-user quality of service (QoS) requirement, successive interference cancellation (SIC) feasibility, and the minimum adjacent antennas spacing constraint. An alternating optimization (AO) approach iteratively solves the proposed non-convex maximization problem. Simulation results demonstrate that the proportional power PASS model achieves higher semantic SE than the benchmark schemes across different user geometries and system-parameter settings.
\end{abstract}
\vspace{-1em}

\vspace{-1em}
\section{Introduction} \label{intro}
Wireless communication has undergone remarkable development since its birth, from enabling high-speed and massive machine-to-machine protocols for the sixth-generation (6G) standard to flexible-antenna systems \cite{ma2023mimo} and intelligent semantic communication (SC) \cite{letaief2019roadmap} to name a few. Pinching-antenna systems (PASS) have emerged as a revolutionary solution to turn the propagation channel into a controllable resource by establishing strong line-of-sight (LoS) links with low-loss dielectric waveguides and cheap clothespins serving as radiating pinching antennas \cite{ding2025flexible}. More specifically, PASS allow non-orthogonal multiple access (NOMA) through these multiple spatially distributed radiating points, which have a significant influence on the effective channels towards the served users \cite{yang2025pinching}.

Existing studies have shown that the locations of the pinching antennas strongly influence the effective PASS channels \cite{xiao2025channel}, \cite{xu2026joint}. These studies have also quantified the effects of waveguide attenuation and antenna spacing in joint transmit-and-pinching beamforming designs. Authors in \cite{zhou2025sum} investigated sum-rate maximization for bit-based communication and derived closed-form expressions for user power allocation and pinching-antenna placement. Beyond single-user bit-rate maximization, PASS have been combined with NOMA through power-domain multiplexing to increase spectral efficiency (SE) \cite{wang2025antenna}.

In this respect, the physics behind the operation of pinching antennas needs to be further studied to support the development of PASS architectures. Most existing studies model PASS similarly to conventional multiple-input multiple-output (MIMO) systems, assuming uniform radiation power across the radiating points \cite{ouyang2025array, wang2025modeling}. Although PASS have started to attract growing research attention, their potential for SC has not been widely explored, particularly for flexible radiation-assisted wireless systems \cite{letaief2019roadmap}, \cite{gunduz2022beyond}, \cite{ahmed2026waveguide}.

State-of-the-art SC systems extract the essential semantic features of a message. The corresponding semantic information rate is expressed in \cite{yan2022resource} as
\vspace{-0.35em}
\begin{equation}
R_{\text{S}} = \frac{WI}{KL} \epsilon_K(\gamma),
\vspace{-0.35em}
\label{semeq}
\end{equation}
where $\epsilon_K(\gamma)$ is the sigmoid-shaped semantic similarity function whose value ranges between zero and one, $\gamma$ represents the received signal-to-noise ratio (SNR), $W$ is the channel bandwidth, $I$ is the amount of semantic information in a message with semantic units (suts), $K$ represents the average number of semantic symbols transmitted for each word, and $L$ denotes the number of words in a sentence. For a unit bandwidth, the resulting $R_{\text{S}}$ is therefore measured in units of suts/s/Hz.

A joint source-channel coding framework, namely DeepSC, provides a practical text-based SC transceiver \cite{xie2021deep}. DeepSC preserves semantic fidelity at low SNRs over fading channels and is implemented using neural networks \cite{getu2025semantic}. In most of the literature, SC is complemented by traditional bit communication as it yields diminishing performance at higher SNR \cite{yan2022resource, ahmed2025semantic}. Therefore, a quality of service (QoS) constraint is commonly imposed for the bit user in a heterogeneous-user network investigation. However, this will require a tailored decoding mechanism due to the neural network architecture for semantic extraction and detection \cite{chen2023uplink, ahmed2025hybrid}. This makes resource allocation particularly important in a network comprising heterogeneous semantic and bit users.

Against this background, we present an adjustable-radiation PASS-assisted framework for heterogeneous semantic and bit users, exploiting the flexibility of pinching antennas and low-power semantic fidelity. More specifically, the main contributions of this work are: i) we formalize a coupling spacing-aware radiation profile into a downlink NOMA-assisted PASS model for heterogeneous users, where the channel coefficients for the semantic and bit users are based on the spherical wavefront model; ii) we adopt a low-complexity alternating optimization (AO) approach to optimize the users power-allocation coefficients and pinching antenna locations, and justify the bit-to-semantic decoding order under the adjustable power PASS to maximize semantic SE while satisfying bit-user QoS; and iii) we provide numerical results to validate the effectiveness of the proportional power PASS against the conventional fixed-antenna systems.

\vspace{-1em}
\section{System Model}
\begin{figure}[t]
\centering
\includegraphics[trim={0cm 0cm 0cm 0cm},clip,width=1\columnwidth]{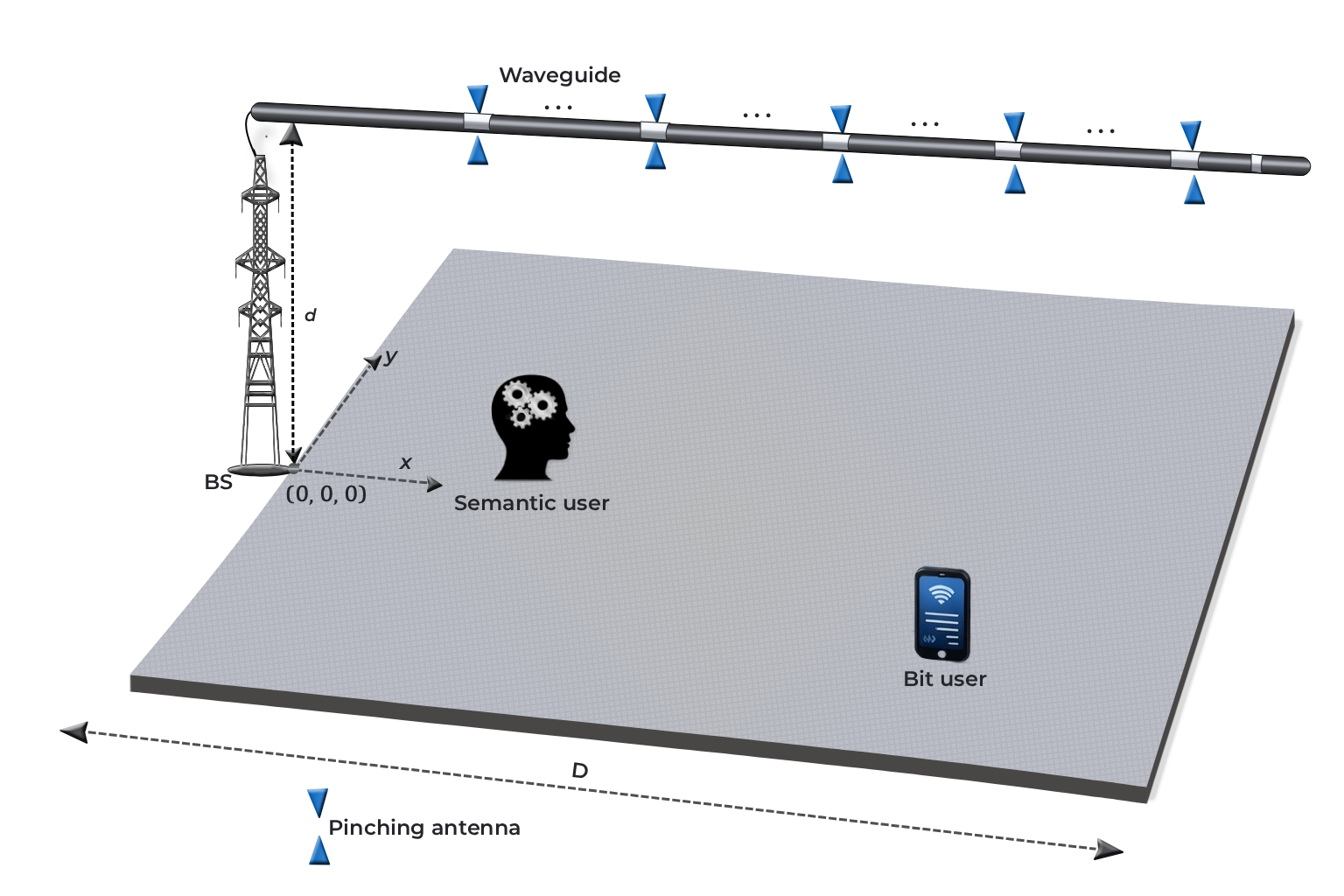}
\caption{An illustration of heterogeneous users NOMA framework via PASS.}
\vspace{-9mm}
\label{fig1}
\end{figure}

We consider a downlink NOMA-assisted framework where a base station (BS) with transmit power $P_{\max }$ serves a semantic user (S) and a bit user (B), via a waveguide mounted at height $d$. As shown in Fig.~\ref{fig1}, the users are randomly located in a square region on the Cartesian plane with side length $D$.
The waveguide is equipped with $N$ unequal-power pinching antennas along its length, such that adjacent antennas are separated by at least $\Delta$, where $\Delta\geq\lambda/2$, and $\lambda$ is the free-space wavelength. Let the $n$-th pinching antenna be at $\tilde{\psi}_n^\mathrm{P}\!=\!(\tilde{x}_n^\mathrm{P}, 0, d)$ with $\tilde{x}_n^\mathrm{P}\!\! \in\![0, D]$, $n \!\in\! N$, and collect their $x$-coordinates in vector $\textbf{x}^\text{P}\!\! =\!\![\tilde{x}_1^\mathrm{P}, \ldots, \tilde{x}_N^\mathrm{P}]$. We denote the radiation amplitude coefficient of the $n$-th pinching antenna by $\beta_n$, and the location of User $m$ by $\phi_m=\left(x_m, y_m, 0\right)$, where $m\! \in\!\{\mathrm{S},\mathrm{B}\}$.
\vspace{-1em}
\subsection{Proportional Power PASS Model}
\begin{figure}[t]
\centering
\includegraphics[trim={0cm 1.5cm 0cm 0cm},clip,width=1\columnwidth]{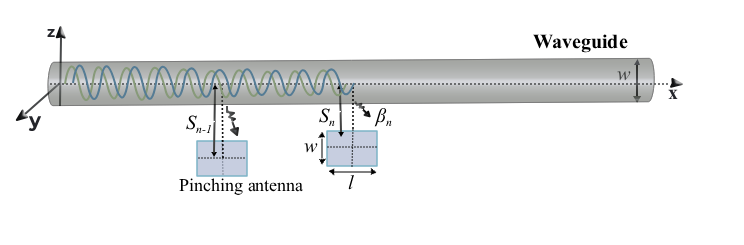}
\caption{Coupling spacing-aware PASS radiation model.}
\vspace{-9mm}
\label{adjpower}
\end{figure}

The proposed framework implements the coupling-aware adjustable power radiation model \cite{wang2025modeling, xu2025pinching}, assuming the same core width \textit{w} for the waveguide and pinching antennas, as shown in Fig.~\ref{adjpower}.
The pinching antenna is essentially modelled as an open-ended directional coupler that couples electromagnetic power from the waveguide into free space, thereby enabling an adjustable $\beta_n$ characterized by coupled-mode theory (CMT) \cite{hermann1991coupled} as
\begin{equation}
\beta_n=\sin \left(\kappa_n l\right)\left(\sqrt{1-\sin ^2\left(\kappa_n l\right)}\right)^{n-1},
\label{radratio}
\end{equation}
where $l$ is the fixed length of each pinching antenna, and $\kappa_n$ denotes the coupling coefficient between the waveguide and the $n$-th pinching antenna \cite{xu2025pinching} expressed as
\vspace{-0.35em}
\begin{equation}
\kappa_n=\Omega_0 e^{-\xi S_n},
\vspace{-0.35em}
\label{kappa}
\end{equation}
where $S_n$ is the coupling spacing, $\Omega_0$ is a coupling constant depending on the physical geometry and dielectric material properties of the coupler formed by the waveguide-antenna, and $\xi$ is the cladding decay constant. For the given power profile, each antenna radiates the same ratio of the remaining power, yielding geometrically decreasing radiated-power fractions.
\vspace{-1em}
\subsection{Signal Propagation Model}
The free-space channel channel coefficient $h_{n,m}$ from the $n$-th pinching antenna to User $m$ is LoS-dominated, given by the spherical wavefront model \cite{zhang2022beam} as
\vspace{-0.35em}
\begin{equation}
h_{n,m}=\frac{\sqrt{\eta} e^{-j \frac{2 \pi}{\lambda}\left|\phi_m-\tilde{\psi}_n^{\mathrm{P}}\right|}}{\left|\phi_m-\tilde{\psi}_n^{\mathrm{P}}\right|},
\vspace{-0.35em}
\label{channel}
\end{equation}
where $\eta=\frac{\lambda^2}{16\pi^2}$ represents the path loss at a reference distance of 1 m, $|\cdot|$ denotes the Euclidean norm, and $j$ is the imaginary unit. Since all the pinching antennas are on the same waveguide and driven by a single RF chain, the BS must superimpose the signals before transmission as
\vspace{-0.35em}
\begin{equation}
s=\sqrt{p_\text{S}} s_\text{S}+\sqrt{p_\text{B}} s_\text{B},
\vspace{-0.35em}
\label{superimposed}
\end{equation}
where $s_\text{S}$ and $s_\text{B}$ are the signals intended for semantic and bit users, respectively, with their power allocation coefficients $p_\text{S}$ and $p_\text{B}$. However, the transmitted signal also includes an additional phase shift $\theta_n$ due to its propagation inside the dielectric waveguide, which lowers its phase velocity relative to free-space. This is captured by the shortened guided wavelength $\lambda_g=\frac{\lambda}{\eta_\text{eff}}$, where $\eta_\text{eff}$ is the effective refractive index of the dielectric waveguide. Consequently, the transmitted signal vector in the adjustable power PASS can be written as
\vspace{-0.5em}
\begin{equation}
\mathbf{s}=\left[\beta_1e^{-j \theta_1}, \cdots, \beta_Ne^{-j \theta_N}\right]^{\mathrm{T}} s,
\vspace{-0.35em}
\label{tx}
\end{equation}
where $\theta_n=2 \pi \frac{\left|\psi_0^{\mathrm{P}}-\tilde{\psi}_n^{\mathrm{P}}\right|}{\lambda_g}$ is the phase shift at the $n$-th pinching antenna, $[\cdot]^{\mathrm{T}}$ denotes the transpose operation, and $\psi_0^{\mathrm{P}}$ is the feed point to waveguide. For the considered system, the received signal at User $m$ is represented as
\vspace{-0.5em}
\begin{equation}
y_m=\mathbf{h}_m^T \mathbf{s}+w_m,
\vspace{-0.35em}
\label{received}
\end{equation}
where $w_m$ denotes the additive white Gaussian noise with variance $\sigma^2$, and \color{black}
\vspace{-0.5em}
\begin{equation}
\mathbf{h}_m=\left[\frac{\sqrt{\eta} e^{-j \frac{2 \pi}{\lambda}\left|\phi_m-\tilde{\psi}_1^{\mathrm{P}}\right|}}{\left|\phi_m-\tilde{\psi}_1^{\mathrm{P}}\right|} \; \cdots \; \frac{\sqrt{\eta} e^{-j \frac{2 \pi}{\lambda}\left|\phi_m-\tilde{\psi}_N^{\mathrm{P}}\right|}}{\left|\phi_m-\tilde{\psi}_N^{\mathrm{P}}\right|}\right]^T.
\vspace{-0.35em}
\label{channelvector}
\end{equation}
The principle of bit-to-semantic decoding is followed for the NOMA-assisted PASS, where the bit user directly decodes its own signal while treating the semantic signal as interference. Therefore, the achievable data rate of User B is given by
\vspace{-0.35em}
\begin{equation}
R_\text{B}^\text{P}\left(\textbf{x}^\text{P}, p_\text{S}\right)=\log _2\left(1+\frac{(1-p_\text{S}) P_{\max } |g_\text{B}|^2}{p_\text{S} P_{\max } |g_\text{B}|^2+\sigma^2}\right),
\vspace{-0.35em}
\label{bitrate}
\end{equation}
where $g_\text{B}\!\!=\!\!\sum\limits_{n \in N}\frac{\sqrt{\eta} e^{-j \frac{2 \pi}{\lambda}|\phi_\text{B}-\tilde{\psi}_n^{\mathrm{P}}|}}{|\phi_\text{B}-\tilde{\psi}_n^{\mathrm{P}}|} \beta_ne^{-j \theta_n}$.
At user S, SIC is first performed to decode and remove the bit-user signal. The achievable rate for decoding the bit-user signal at User S is given by
\vspace{-0.35em}
\begin{equation}
R_{\text{B}\rightarrow \text{S}}^\text{P}\left(\textbf{x}^\text{P}, p_\text{S}\right)=\log _2\left(1+\frac{(1-p_\text{S}) P_{\max } |g_\text{S}|^2}{p_\text{S} P_{\max } |g_\text{S}|^2+\sigma^2}\right),
\vspace{-0.35em}
\label{SICeqrate}
\end{equation}
where $g_\text{S}\!\!=\!\!\sum\limits_{n \in N}\frac{\sqrt{\eta} e^{-j \frac{2 \pi}{\lambda}\left|\phi_\text{S}-\tilde{\psi}_n^{\mathrm{P}}\right|}}{\left|\phi_\text{S}-\tilde{\psi}_n^{\mathrm{P}}\right|} \beta_ne^{-j \theta_n}$. The pinching-antenna positions are optimized to strengthen the LoS channel of User S while satisfying the specified minimum rate $R_{\text{B} }^{\text{min}}$ of User B. Subsequently, User S decodes its signal in an interference-free manner for a unit bandwidth as
\vspace{-0.35em}
\begin{equation}
R_\text{S}^\text{P}\left(\textbf{x}^\text{P}, p_\text{S}\right)=\frac{I}{K L} \epsilon_K(\gamma_\text{S})
\vspace{-0.35em}
\label{semanticrate},
\end{equation}
where $\gamma_\text{S}=\frac{p_\text{S} P_{\max } |g_\text{S}|^2}{\sigma^2}$. In practice, the closed-form expression of $\epsilon_K(\gamma_\text{S})$ is not available, so the generalized logistic approximation is adopted via data regression on DeepSC outputs. Specifically, for each $K$ the DeepSC tool is run over a grid of $\gamma_\text{S}$ values to obtain empirical $\epsilon_K(\gamma_\text{S})$ samples. By evaluating DeepSC for different values of $K$ and $\gamma_\text{S}$, $\epsilon_K(\gamma_\text{S})$ is observed to be monotonically non-decreasing in $\gamma_\text{S}$ \cite{yan2022resource}. Its slope initially increases and subsequently decreases, producing a sigmoid-shaped pattern bounded between zero and one. Authors in \cite{mu2022heterogeneous} deployed a data-regression method to tractably approximate the values of $\epsilon_K(\gamma_\text{S})$ by following the criterion of minimum mean square error for fitting the values with a generalized logistic function as expressed by
\vspace{-0.35em}
\begin{equation}
\epsilon_K(\gamma_\text{S}) \overset{\triangle}{=} A_{K,1} + 
\frac{A_{K,2} - A_{K,1}}{1 + e^{-(C_{K,1} \gamma_\text{S} + C_{K,2})}},
\vspace{-0.35em}
\label{logisticapprox}
\end{equation}
where the lower (left) asymptote, upper (right) asymptote, growth rate, and the mid-point parameters of the logistic function are respectively denoted by $A_{K,1}$, $A_{K,2}$, $C_{K,1}$, and $C_{K,2}$ for different values of $K$.

\vspace{-1em}
\section{Problem Formulation}
Our objective is to maximize the SE for User S while guaranteeing the QoS for User B, and invoking the bit-to-semantic decoding order. An optimization problem is formulated so that the rates for both users depend jointly on $\textbf{x}^\text{P}$ and $p_\text{S}$, as given by
\vspace{-0.35em}
\begin{align}
\mathbf{(P0)}:\max _{\textbf{x}^\text{P}, p_\text{S}} \quad & R_\text{S}^\text{P}\left(\textbf{x}^\text{P}, p_\text{S}\right), \label{originalobjfunc}\\
\text { s.t. } \quad & \left|\tilde{x}_n^P-\tilde{x}_{n-1}^P\right| \geq \Delta, \! \forall n \in\{2, \ldots, N\}, \label{antennaspacing}\\
& R_\text{B}^\text{P}\left(\textbf{x}^\text{P}, p_\text{S}\right) \geq R_{\text{B} }^{\text{min}}, \label{bituserQoS}\\
& R_{\text{B}\rightarrow \text{S}}^\text{P}\left(\textbf{x}^\text{P}, p_\text{S}\right) \geq R_{\text{B} }^{\text{min}}, \label{SIC}\\
& 0\leq p_\text{S}\leq p_\text{B}, \label{feasiblePA}\\
& p_\text{S}+p_\text{B}=1. \label{ratiosum}
\end{align}
Constraint \eqref{antennaspacing} enforces the minimum adjacent antennas spacing to prevent inter-channel coupling, while \eqref{bituserQoS} and \eqref{SIC} ensure bit-user QoS and SIC feasibility under the bit-to-semantic decoding order prescribed by \eqref{feasiblePA}. Constraint \eqref{ratiosum} guarantees that the available transmit power is distributed between the semantic and bit users. The formulated maximization problem is non-convex as $\epsilon_K(\gamma_\text{S})$ fails to satisfy concavity in $\gamma_\text{S}$.

\vspace{-1em}
\subsection{Solution Method}
We decouple the problem via AO into two subproblems and solve as follows.

\vspace{-1em}
\subsection{Power Allocation Subproblem}
In this subproblem, $\textbf{x}^\text{P}$ is assumed to be fixed and feasible, which simplifies $\mathbf{(P0)}$ as
\vspace{-0.35em}
\begin{align}
\max _{p_\text{S}} \quad & R_\text{S}^\text{P}\left( p_\text{S}\right), \label{PAobjfunc}\\
\text { s.t. } \quad
& R_\text{B}^\text{P}\left(p_\text{S}\right) \geq R_{\text{B} }^{\text{min}}, \label{PAbituserQoS}\\
& R_{\text{B}\rightarrow \text{S}}^\text{P}\left(p_\text{S}\right) \geq R_{\text{B} }^{\text{min}}, \label{PASIC}\\
& 0\leq p_\text{S}\leq p_\text{B}, \label{subfeasiblePA}\\
& p_\text{S}+p_\text{B}=1.
\end{align}
It is important to consider that the above simplified problem is still non-convex due to the dependence on $\epsilon_K(\gamma_\text{S})$. However, since $\epsilon_K(\gamma_\text{S})$ is monotonically non-decreasing in $\gamma_\text{S}$ for the given antenna positions, the optimal $p_\text{S}$ is the largest value satisfying the bit-user QoS, SIC feasibility, and NOMA decoding constraints.

The optimal power allocation is decided on the basis of active constraints for the sigmoid-shaped bounded objective function. From \eqref{PAbituserQoS}, algebraic manipulations yield the closed-form solution as
\vspace{-0.35em}
\begin{equation}
p_\text{S} \leq \frac{P_{\max } h_{\text{B}}-\tau \sigma^2}{P_{\max }h_{\text{B}}(1+\tau)},
\vspace{-0.35em}
\label{alphas}
\end{equation}
where $\tau=2^{R_{\text{B} }^{\text{min}}}-1$, and $P_{\max } h_{\text{B}}-\tau \sigma^2 \geq 0$ for feasibility at the given $\textbf{x}^\text{P}$. Likewise, the SIC decodability constraint results in the following closed-form upper bound $p_\text{S-SIC}$ on the semantic-user power allocation coefficient.
\vspace{-0.35em}
\begin{equation}
p_\text{S-SIC} \leq \frac{P_{\max } h_{\text{S}}-\tau \sigma^2}{P_{\max }h_{\text{S}}(1+\tau)},
\vspace{-0.35em}
\label{alphasic}
\end{equation}
Following the proof in \cite{fu2025power}, the optimal power coefficient is obtained when the closed-form solutions hold as equalities. The optimized power allocation coefficient $p_\text{S}^*$ can therefore be written as
\vspace{-0.35em}
\begin{equation}
p_\text{S}^*=\max \left\{0, \min \left\{p_\text{S}, p_\text{S-SIC}, 0.5\right\}\right\}.
\vspace{-0.35em}
\label{alphas*}
\end{equation}

\vspace{-1em}
\subsection{Antennas Position Subproblem}
In this subproblem, our aim is to strategically determine the deployment of pinching antennas based on a fixed $p_\text{S}^*$ value. The pinching antenna position affects the channel through two coupled mechanisms, namely the distance-dependent spherical-wave free-space component and the guided-wave phase component along the dielectric waveguide. Therefore, the optimal value of $\textbf{x}^\text{P*}$ that maximizes the semantic SE in \eqref{originalobjfunc} is of great importance. Moreover, to cope with the phase shifts due to the propagation along the waveguide, it is necessary to fine-tune the pinching-antenna deployment to ensure phase alignment of the signal through free-space following waveguide propagation. Therefore, this subproblem involves two coordinated steps, namely large-scale antenna placement and fine-scale phase alignment. In the first step, we iteratively relocate the pinching antennas on the dielectric waveguide to form the large-scale channel and enhance the effective channel gain of the semantic user while maintaining the bit-user QoS. This is implemented via a one-dimensional bisection search such that the first pinching antenna is placed where the User S experiences the best channel gain, with the remaining antennas evenly placed and satisfying the minimum-spacing constraint. Subsequently, small phase adjustments are performed to enhance the composite channel gain of the semantic user.

Let us assume that the pinching antennas are placed sequentially along the x-axis on the waveguide, with adjacent ones satisfying the minimum-spacing condition. Based on this, the antenna position subproblem is formulated as
\vspace{-0.35em}
\begin{align}
\max _{\textbf{x}^\text{P}} \quad & R_\text{S}^\text{P}\left( \textbf{x}^\text{P}\right), \label{APobjfunc}\\
\text { s.t. } \quad & \left|\tilde{x}_n^P-\tilde{x}_{n-1}^P\right| \geq \Delta, \! \forall n \in\{2, \ldots, N\}, \label{APspacing}\\
& R_\text{B}^\text{P}\left(\textbf{x}^\text{P}\right) \geq R_{\text{B} }^{\text{min}}, \label{APbituserQoS}\\
& R_{\text{B}\rightarrow \text{S}}^\text{P}\left(\textbf{x}^\text{P}\right) \geq R_{\text{B} }^{\text{min}}, \label{APSIC}\\
& \left|\phi_{\text{S}, n}-\phi_{\text{S}, n-1} \pm 2\pi i\right| \leq \delta_\text{S}, \! \forall n \in\{2, \ldots, N\}, \label{phaseS}\\
& \left|\phi_{\text{B}, n}-\phi_{\text{B}, n-1} \pm 2\pi i\right| \leq \delta_\text{B}, \! \forall n \in\{2, \ldots, N\} \label{phaseSB}
\end{align}
here $\phi_{m,n}\!=2\pi \left( \frac{|\phi_m - \tilde{\psi}_n^{\mathrm{P}}|}{\lambda} - \frac{|\psi_0^{\mathrm{P}} - \tilde{\psi}_n^{\mathrm{P}}|}{\lambda_g} \right)$, such that $m\! \in\!\{\mathrm{S},\mathrm{B}\}$ and includes the free-space and waveguide propagation distance terms. Constraints \eqref{phaseS} and \eqref{phaseSB} ensure phase alignment for the semantic and bit users, respectively. An arbitrary integer $i$ accounts for the $2 \pi$ phase periodicity, while $\delta_\text{S}$ and $\delta_\text{B}$ represent the predefined positive phase-precision constants for these users.

For any User $m$, the adjacent antennas phase difference is computed as $\Delta \phi_{m, n}\!=\!\!\phi_{m, n}-\phi_{m, n-1}$. Since phase is periodic, the nearest integer multiple of $2 \pi$ is removed. Notably, the phase-precision conditions are not assumed to be always jointly satisfiable for every channel realization. Instead, they are used as feasibility and phase-refinement criteria within the position update step. Specifically, after the large-scale bisection-based antennas placement, the phase fine-tuning step $\tilde{\Delta}$ reduces the residual phase mismatch. A fine-tuned candidate is accepted only if it satisfies the minimum-spacing constraint, the semantic-user and bit-user phase-precision conditions, the bit-user QoS constraint, and the SIC feasibility constraint. Otherwise, the candidate is discarded and the previous feasible pinching antennas placement is retained. If no feasible placement exists for a given channel realization, that realization is treated as infeasible under the considered constraints. Therefore, the accepted pinching antennas-position sequence is non-decreasing by construction despite the heuristic search-based approach.

For each channel realization, the AO procedure iteratively updates $p_\text{S}$ and $\textbf{x}^\text{P}$ such that the resulting sequence of semantic SE values is monotonically non-decreasing and bounded. The power-allocation update maximizes the semantic SE over the feasible interval, and has a constant complexity, i.e., $\mathcal{O}(1)$. For the antennas position optimization, each bisection step involves the contributions of all $N$ pinching antennas and therefore requires $\mathcal{O}(N)$ operations. Reducing an initial search interval of length $M$ to an accuracy $\varepsilon$ requires $\mathcal{O}(\log_2(M/\varepsilon))$ bisection iterations, while the stage of phase refinement scales linearly with $N$. Therefore, the overall computational complexity is $\mathcal{O}(N\log _2(M / \varepsilon))$.

\vspace{-1em}
\section{Simulation Results}
For performance evaluation of the NOMA-assisted semantic enhanced PASS model, we use the following simulation parameters: $\sigma^2=-90$\! dBm, the carrier frequency $f_c\!=\!28$\! GHz, $d\!=\!3$\! m, \textit{w} \!=\! 10\! mm, $\eta_\text{eff}=\!1.4$, $\Delta\!=\!\lambda/2$, $\tilde{\Delta}\!=\!\lambda/10$, $R_{\text{B} }^{\text{min}}\!=\!0.5$\! bps/Hz, $D\!\in\!\{20,40\}$\! m, and $N\!\in\!\!\{3,7\}$. For SC, the parametric values of $K\!\!=\!\!5$, $I/L=1$, $A_{K,1}\!=\!0.37$, $A_{K,2}\!=\!0.98$, $C_{K,1}\!=\!0.25$, and $C_{K,2}\!=\!-0.7895$ are used \cite{mu2023exploiting}. Following \cite{xu2025pinching}, the waveguide coupling parameters are calculated using $\xi=0.24615$\! mm$^{-1}$, $\Omega_0=0.3300$\! mm$^{-1}$, and $l=5$\! mm, with a minimum coupling spacing $S_n$ of $0.2$ mm. The total radiated power ratio across all antennas satisfies $\sum\limits_{n \in N} \!\beta_n^2\! \leq 1$. For a given $N$, $S_n$ is identical across all pinching antennas and remains fixed throughout the AO procedure. As benchmark schemes, we consider a NOMA-assisted conventional antenna system (CAS) and an equal power PASS where all antennas radiate equally \cite{wang2025modeling}. For the CAS, the BS is centered within the service region at height $d$ and equipped with an equivalent number of fixed antennas at half-wavelength spacing \cite{ding2025flexible}. Moreover, the semantic SE is averaged over $10^5$ realizations.

\begin{figure}[!t]
\centering
\includegraphics[trim={0cm 0cm 0cm 0cm},clip,width=1\columnwidth]{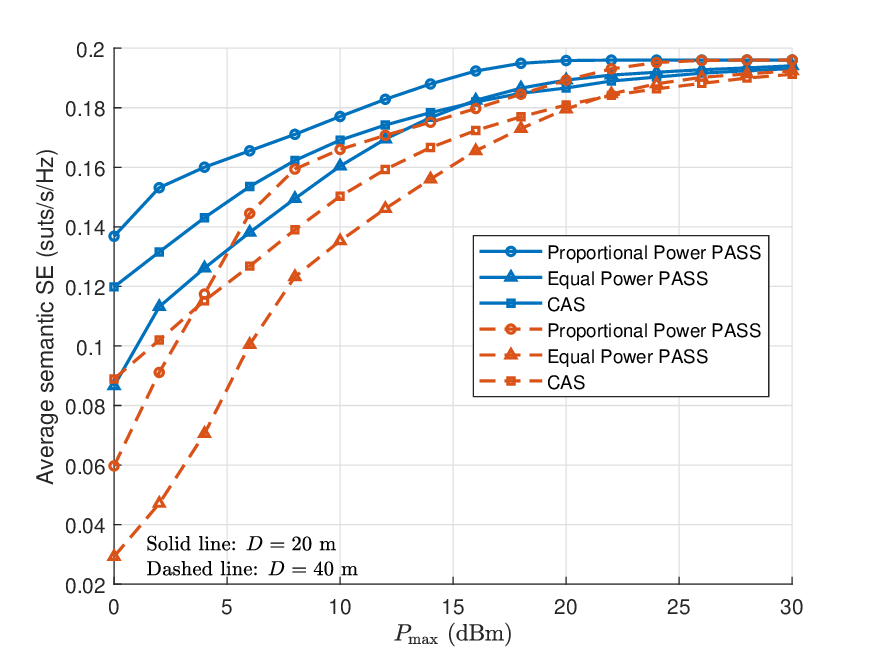}
\caption{Average semantic SE versus $P_{\max }$ for the NOMA-assisted adjustable power PASS and CAS with $N\!=\!3$.}
\vspace{-7mm}
\label{fig2}
\end{figure}

Fig.~\ref{fig2} illustrates the average semantic SE versus $P_{\max}$ for the adjustable power PASS and CAS models. Notably, the equal power PASS refers to the case of adjustable PASS in which the coupling spacing between the waveguide and pinching antennas is sequentially adjusted to emit equal power from each of the $N$ radiating points \cite{xu2025pinching}. As $P_{\max}$ increases, the average semantic SE improves quickly before becoming more gradual. Evidently, the proportional power PASS achieves a higher semantic SE than the equal power PASS and CAS, with performance determined by the radiation model and reconfigurable propagation channels. Furthermore, the performance improves for a smaller coverage area because the corresponding propagation distances are shorter.

\begin{figure}[!t]
\centering
\includegraphics[trim={0cm 0cm 0cm 0cm},clip,width=1\columnwidth]{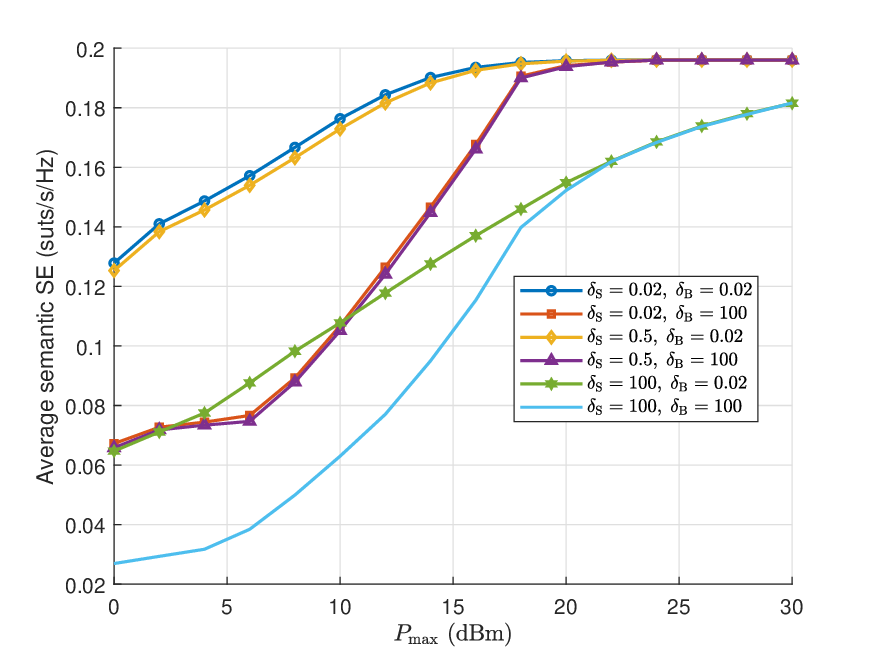}
\caption{Average semantic SE versus $P_{\max }$ under different phase alignments for the proportional power PASS with $N\!=\!3$ and $D\!=\!20$\! m.}
\vspace{-9mm}
\label{fig3}
\end{figure}

Fig.~\ref{fig3} shows the average semantic SE versus $P_{\max}$ under different phase alignment accuracies for $\delta_\text{S}$ and $\delta_\text{B}$ for the proportional power PASS with $N=3$ and $D=20$ m. The values 0.02, 0.5, and 100 radians for $\delta_\text{S}$ and $\delta_\text{B}$ can be interpreted as fine, moderate, and coarse phase alignment, respectively. More specifically, constraining $\delta_\text{S}$ with a smaller value offers the best performance, as it directly boosts the objective function involving semantic SE. With $\delta_\text{S}=0.02$ and $\delta_\text{B}=100$, the curve trails the best-performing curve at low $P_{\max}$, but its performance closely matches the top curve from 18 dBm onwards. By contrast, with $\delta_\text{S}=100$ and $\delta_\text{B}=0.02$, performance degrades noticeably because the coarse $\delta_\text{S}$ suppresses the semantic user's performance. When both users are coarsely aligned, the semantic SE curve remains at the lowest level across all $P_{\max}$ values.

\begin{figure}[!t]
\centering
\includegraphics[trim={0cm 0cm 0cm 0cm},clip,width=1\columnwidth]{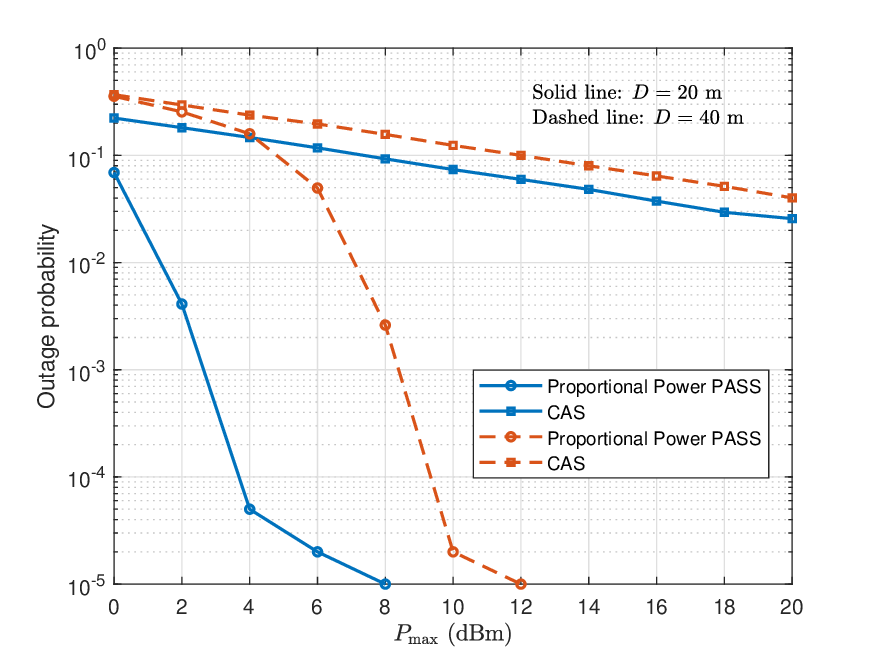}
\caption{Outage probability of the bit-user QoS and SIC feasibility versus $P_{\max}$ for the proportional power PASS and CAS with $N\!=\!3$.}
\vspace{-7mm}
\label{fig4}
\end{figure}

Fig.~\ref{fig4} compares the probability that either the bit-user QoS constraint or the SIC-feasibility constraint is violated for proportional power PASS and CAS. The former consistently yields a lower outage due to its flexibility in repositioning pinching antennas with radiation control capability. In contrast, due to the fixed antenna positions in CAS, random geometries are more likely to violate feasibility conditions. Furthermore, a smaller coverage area results in a lower outage probability because the corresponding propagation links are shorter. As $P_{\max}$ increases, the curves decay monotonically, where the proportional PASS attains negligible outage probability at lower power levels. 

\begin{figure}[!t]
\centering
\includegraphics[trim={0cm 0cm 0cm 0cm},clip,width=1\columnwidth]{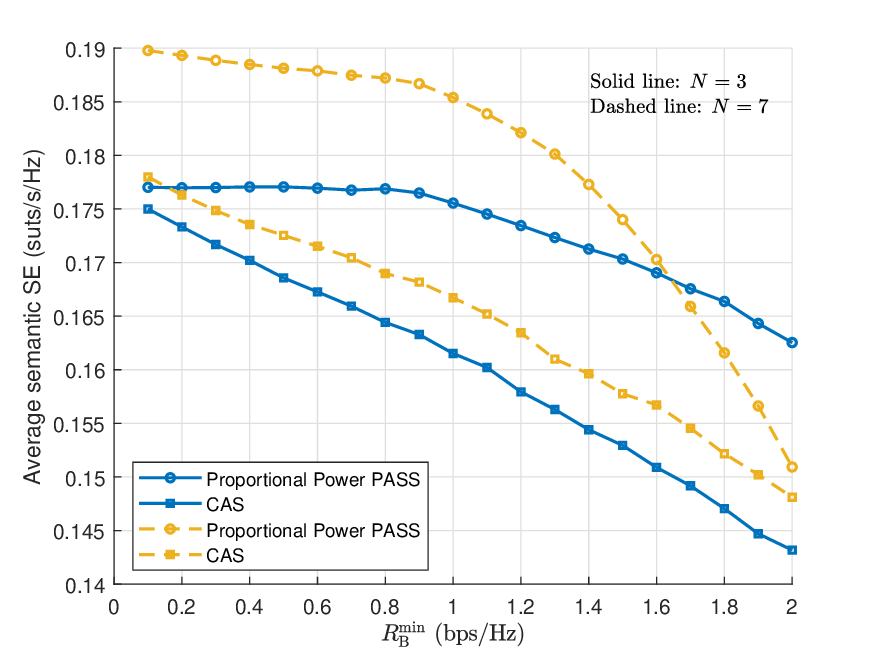}
\caption{Average semantic SE versus $R_{\text{B} }^{\text{min}}$ for the proportional power PASS and CAS at $P_{\max}\!=\!10$ dBm and $D\!=\!20$\! m.}
\vspace{-9mm}
\label{fig5}
\end{figure}

Fig.~\ref{fig5} illustrates the average semantic SE performance for the compared proportional PASS and CAS with varying bit-user QoS requirement. The semantic SE gradually decreases as $R_{\text{B}}^{\text{min}}$ increases because a larger fraction of the available transmit power must be allocated to the bit user. However, proportional PASS consistently achieves better performance across the entire range. Moreover, for $N\!=\!7$, the proportional radiation profile is distributed across more antennas than for $N\!=\!3$ case. This reduces the dominance of the strongest radiating elements, can degrade feasibility as $R_{\text{B} }^{\text{min}}$ increases.

\begin{figure}[!t]
\centering
\includegraphics[trim={0cm 0cm 0cm 0cm},clip,width=1\columnwidth]{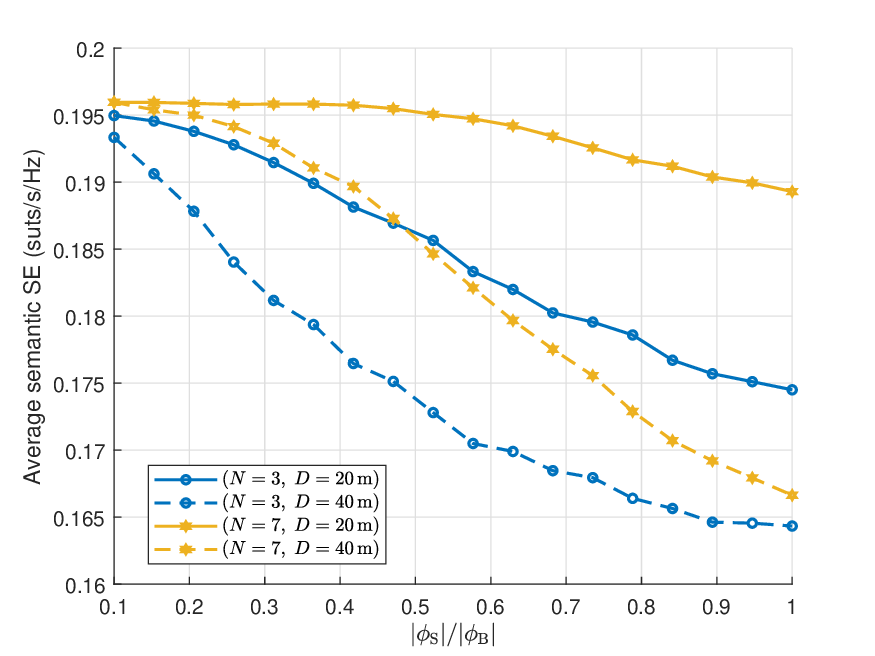}
\caption{Average semantic SE versus users distance ratio from the origin for the proportional power PASS at $P_{\max}\!=\!10$ dBm.}
\vspace{-7mm}
\label{fig6}
\end{figure}

Fig.~\ref{fig6} shows the average semantic SE versus the ratio of the semantic-user distance from the origin to the bit-user distance from the origin for the proportional PASS. Notably, the averaging is done by grouping all realizations of the ratio within uniform ratio intervals. As the ratio increases, the semantic and bit users become located at comparable distances from the origin such that the value 1 implies that both are at equal Euclidean distances. It is evident that the performance gain is more pronounced in the higher distance ratio regimes for smaller $D$. 

\begin{figure}[!t]
\centering
\includegraphics[trim={0cm 0cm 0cm 0cm},clip,width=1\columnwidth]{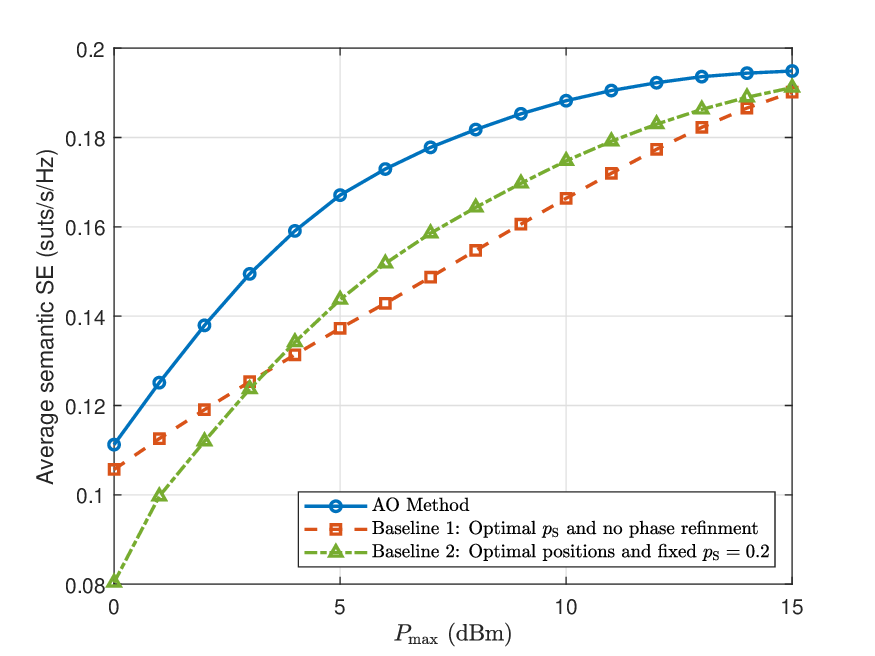}
\caption{Comparison of the AO strategy for the proportional power PASS with baselines using $N=7$ and $D=20$ m.}
\vspace{-9mm}
\label{fig7}
\end{figure}

Fig.~\ref{fig7} presents an ablation study of the proposed AO method for heterogeneous users power allocation and pinching antenna positions with $N\!=\!7$ and $D\!=\!20$\! m. The gain over Baseline 1 demonstrates the benefit of phase refinement in pinching-antenna position update, whereas the more pronounced gap relative to Baseline 2 at low $P_{\max}$ confirms the importance of proposed power allocation strategy under the given QoS and SIC constraints.

\vspace{-1em}
\section{Conclusions}
This paper investigated an adjustable power PASS framework in which multiple pinching antennas are iteratively positioned along a waveguide to serve heterogeneous semantic and bit users through NOMA. An iterative algorithm is proposed for the optimization of user power-allocation coefficients and pinching-antenna positions for maximizing the semantic SE. Simulation results validate the approach by showing clear gains over the conventional baseline. A promising research direction is to jointly optimize the antenna-waveguide coupling spacings, pinching-antenna positions, and user power-allocation coefficients for multiuser heterogeneous semantic and bit communication.


\bibliographystyle{IEEEtran}
\color{black}
\bibliography{references}
\end{document}